\documentclass[a4paper,11pt]{article}
\usepackage{pos}
\usepackage{soul}
\usepackage{lineno}
\usepackage{xcolor}

\title{A Search for Neutrinos from Decaying Dark Matter in Galaxy Clusters and Galaxies with IceCube}
 \ShortTitle{Search for extragalactic DM with IceCube}

\author{The IceCube Collaboration \\{\normalsize \normalfont(a complete list of authors can be found at the end of the proceedings)}}

\emailAdd{minjin.jeong@g.skku.edu}

\abstract{The observed dark matter abundance in the Universe can be explained with non-thermal, heavy dark matter models. In order for dark matter to still be present today, its lifetime has to far exceed the age of the Universe. In these scenarios, dark matter decay can produce highly energetic neutrinos, along with other Standard Model particles. To date, the IceCube Neutrino Observatory is the world’s largest neutrino telescope, located at the geographic South Pole. In 2013, the IceCube collaboration reported the first observation of high-energy astrophysical neutrinos. Since then, IceCube has collected a large amount of astrophysical neutrino data with energies up to tens of PeV, allowing us to probe the heavy dark matter models using neutrinos. We search the IceCube data for neutrinos from decaying dark matter in galaxy clusters and galaxies. The targeted dark matter masses range from 10\,TeV to 10\,PeV. In this contribution, we present the method and sensitivities of the analysis.

\vspace{4mm}
{\bfseries Corresponding authors:}
M. Jeong$^{1*}$\\
{$^{1}$ \itshape Sungkyunkwan University}\\[4mm]
$^*$ Presenter

\FullConference{37$^{\rm{th}}$ International Cosmic Ray Conference (ICRC 2021)\\
		July 12th -- 23rd, 2021\\
		Online -- Berlin, Germany}

}
\begin{document}
\maketitle
\section{Introduction}

\indent The existence of dark matter has been well established by a variety of astronomical observations, but the particle nature of dark matter still remains unknown. Various theories, beyond the Standard Model of particle physics, propose candidates for dark matter. Among those, Weakly Interacting Massive Particles (WIMPs) have been the most extensively studied candidates, for a number of reasons. WIMPs naturally explain the abundance of dark matter in the present Universe~\cite{Steigman:2012nb}. On the other hand, the mass scale allowed for WIMPs is limited up to $\sim$100\,TeV, due to the unitarity bound~\cite{unitarity_bound}, and thus  accessible by current instruments. However, decaying heavy dark matter, which was non-thermally produced in the early Universe and has a longer lifetime than the age of the Universe, is also a viable candidate for dark matter~\cite{Murase:2012xs,Rott:SDM}.  

\indent Neutrinos are useful tools for testing dark matter hypotheses. When dark matter pair-annihilates or decays, neutrinos could be produced directly or through the decay of the primary products. Hence, the properties of dark matter can be constrained by measuring the neutrino flux from celestial objects which host a large amount of dark matter. Neutrinos have several advantages over other messenger particles. They are not affected by magnetic fields and have extremely small cross-sections. Thus, their directions can be better associated with their origins. Also, they are barely absorbed at their production site. This provides unique opportunities to search for dark matter in the core of the Sun and the Earth~\cite{IceCube:SolarWIMP_3yr,IceCube:EarthWIMPs_EPJC17,ANTARES:SecludedDM}. 

\indent The IceCube Neutrino Observatory~\cite{IceCube:detector} is a cubic-kilometer scale neutrino telescope deployed in the deep glacial ice in the Antarctica. The IceCube collaboration has reported observations of high-energy astrophysical neutrino events~\cite{IceCube:science2013,IceCube:HESE_3yr_PRL,IceCube:AstroNuMu_PRL}. Theoretical studies suggest that some of these astrophysical neutrinos could originate from the decay of dark matter particles~\cite{Esmaili:2014rma,Boucenna:2015tra,Chianese:2017nwe,Bhattacharya:2019ucd}. Recent IceCube analyses looked for neutrinos from dark matter decay in the Galactic Halo and isotropic neutrinos from extragalactic dark matter decay~\cite{IceCube:GH,IceCube:HESE_DM_ICRC2019}. These searches provide highly competitive limits on the dark matter decay lifetime, demonstrating that IceCube is sensitive to probe decaying heavy dark matter hypotheses.

\indent In this contribution, we present a search for neutrinos from the decay of dark matter particles in galaxy clusters and galaxies, which utilizes track-like events observed in IceCube. In section~\ref{sec:2}, we describe the expected signal and background of this analysis. In section~\ref{sec:3}, we discuss the data sample and statistical methods used in this work. Then we present the sensitivities of this analysis in section~\ref{sec:4} and conclude in section~\ref{sec:5}.

\section{Signal and Background} \label{sec:2}
\begin{figure}[b]
\centering
\includegraphics[width=.48\linewidth]{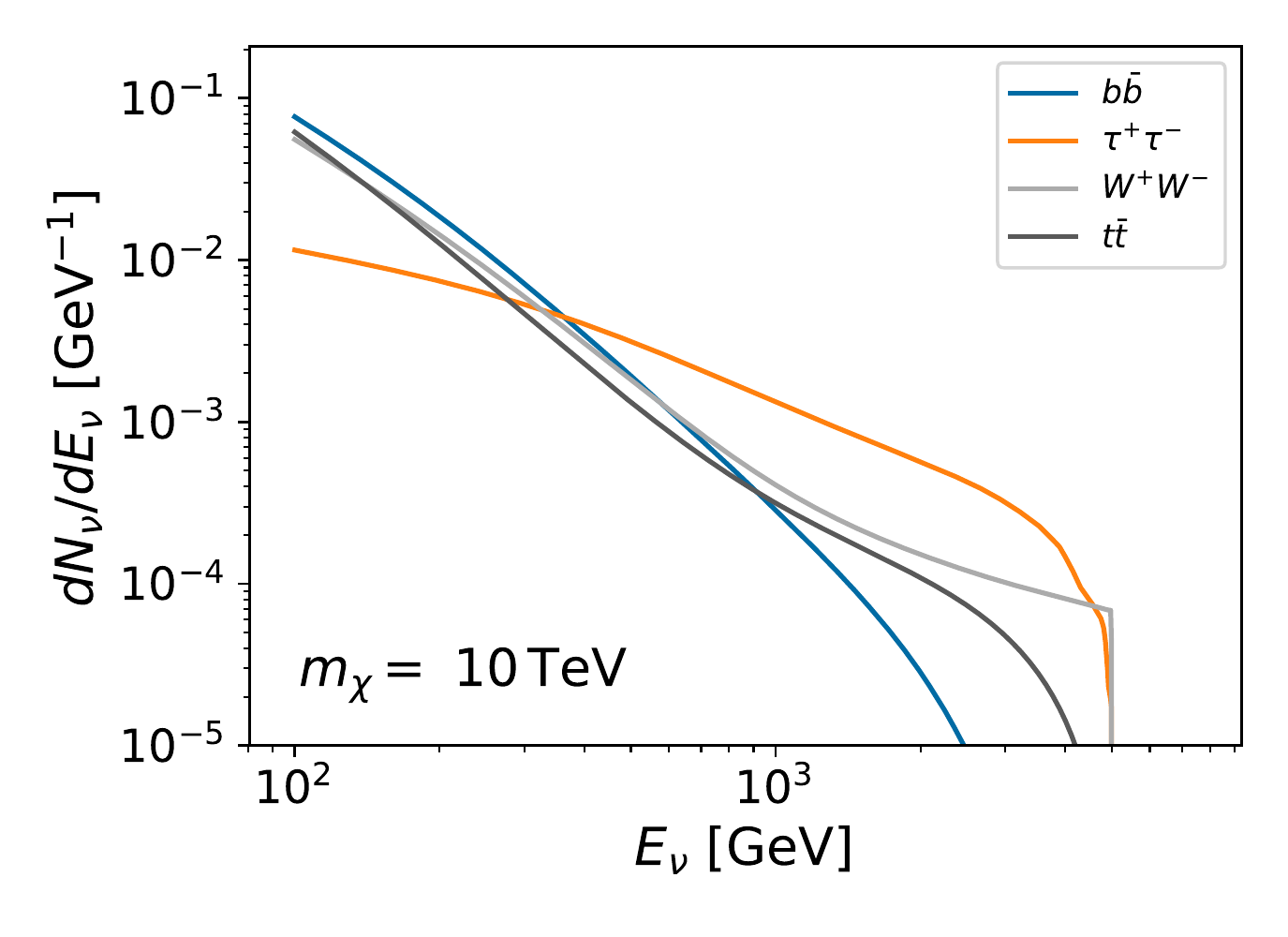}  
\includegraphics[width=.48\linewidth]{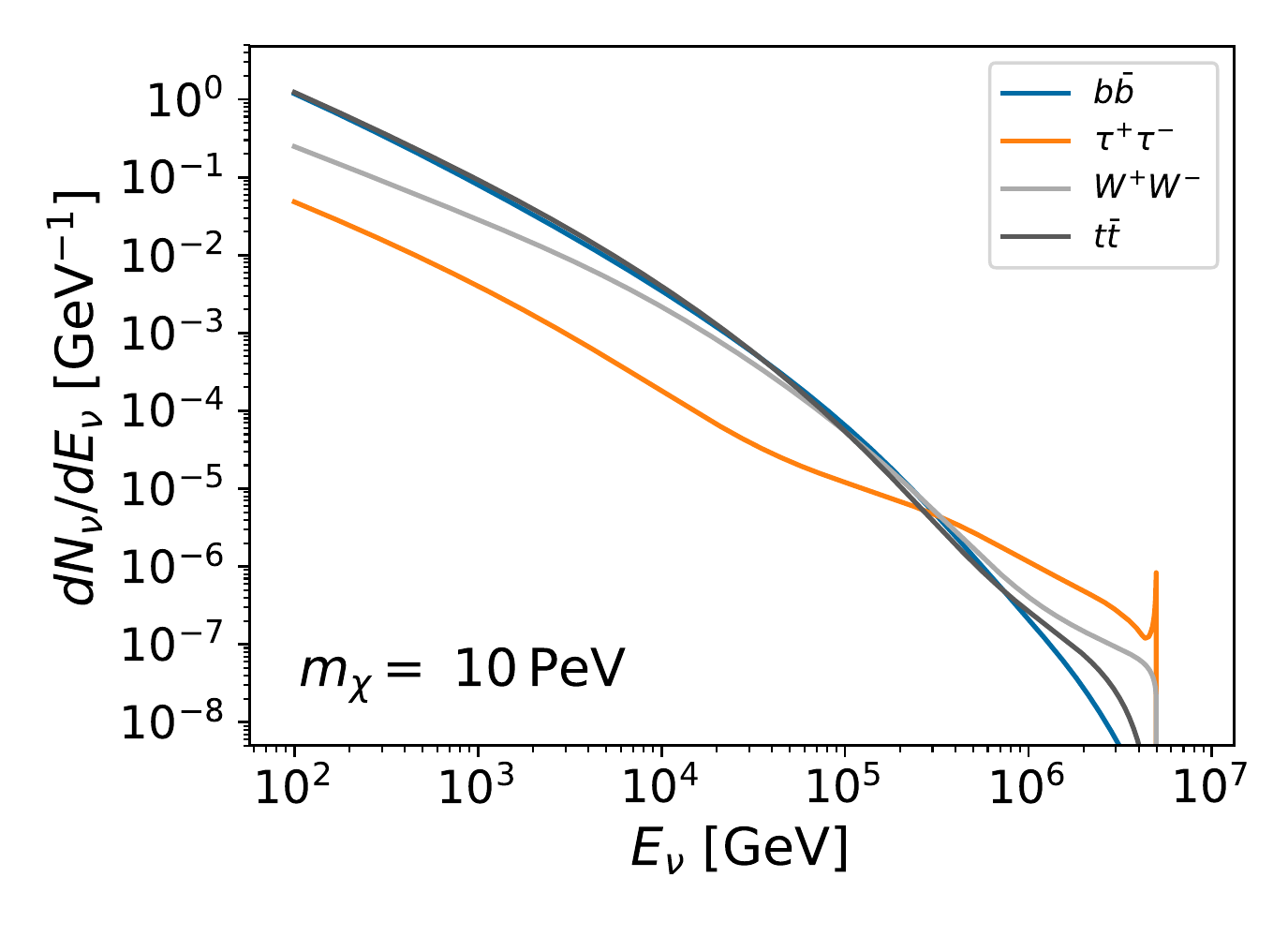}  
\caption[margin=1cm]{Differential energy spectra of $\nu_{\mu}$ (and $\bar{\nu}_{\mu}$) expected at the Earth for four different dark matter decay channels and two different dark matter masses.}
\label{fig:neutrino_spectra}
\end{figure}

The differential neutrino flux expected from the decay of dark matter particles in an extragalactic source is calculated as the following:
\begin{equation}
\frac{d\Phi_{\nu}}{dE_{\nu}}(E_{\nu}) 
    = \frac{1}{4\pi{m}_{\chi}{\tau}_{\chi}} \frac{dN_{\nu}}{dE_{\nu}}(E_{\nu}) 
    \int^{\Delta \Omega}_{0} d\Omega \int^{\infty}_{0}\rho_{\chi}(l\hat{n})dl,
\label{eq:flux}
\end{equation}
with $m_{\chi}$ being the dark matter mass, $\tau_{\chi}$ the dark matter lifetime, $\hat{n}$ the direction of the line-of-sight, and $d{N}_{\nu}/d{E}_{\nu}$ the differential neutrino spectrum per dark matter particle decay. We assume that dark matter decays into a pair of Standard Model particles with a branching ratio of 100\% and calculate the energy spectrum of neutrinos in the final state. This calculation is done for different dark matter masses and decay channels using the $\chi aro\nu$ software package~\cite{Charon}. This package takes into account the electroweak corrections and the neutrino oscillation from the source to the Earth. Figure~\ref{fig:neutrino_spectra} shows the expected spectra of muon neutrinos (and anti-muon neutrinos) at the Earth, for dark matter masses of 10\,TeV and 10\,PeV, decaying through different dark matter channels. The neutrino oscillation inside the Earth is negligible, as we use neutrinos with energies higher than 100\,GeV. The attenuation of high-energy neutrino fluxes due to absorption in the Earth is accounted for using the Nugen software packages~\cite{Nugen}.

\indent The spatial distribution of the neutrino flux depends on the distribution of dark matter in the source. In this analysis, we use galaxy clusters, the Andromeda galaxy (M31), and satellite dwarf galaxies of the Milky Way as targets. For dwarf galaxies, we adopt the dark matter halo models presented in ~\cite{Geringer-Sameth:2014}. The authors use the Zhao density profile~\cite{Zhao:1995cp} to describe the dark matter halos in the dwarf galaxies. For galaxy clusters, we adopt the models presented in~\cite{SanchezConde:2011ap}. Here, the dark matter halos of these clusters are described using the Navarro-Frenk-White (NFW) profile~\cite{NFW:1996} which is a special case of the Zhao profile. The model for the dark matter distribution in the Andromeda galaxy is adopted from~\cite{Tamm:2012hw} and is based on the Einasto profile~\cite{Graham:Einasto_profile_2005}. Lastly, we assume that the dark matter halos of the sources do not contain subhalos, as the signal from decaying dark matter is not highly sensitive to subhalos.  

\begin{figure}[b]
\centering
\includegraphics[width=.8\linewidth]{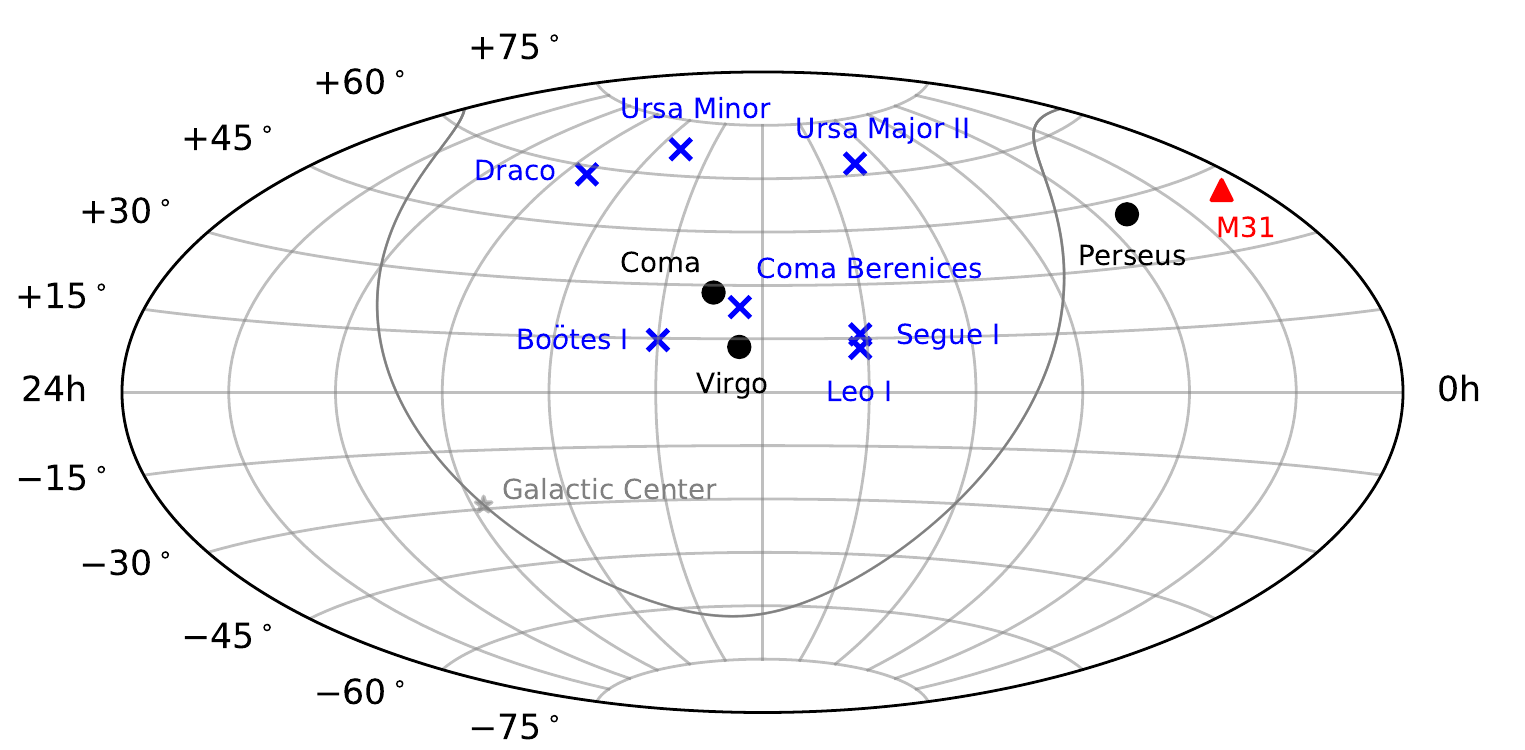}  
\caption[margin=1cm]{Locations of the sources in equatorial coordinates. The circles denote the location of the galaxy clusters. The crosses are for the dwarf galaxies, and the triangle is for the Andromeda galaxy (M31). For references, the locations of the Galactic Center (star) and the Galactic Plane (line) are also shown.}
\label{fig:skymap}
\end{figure}

\indent Given the dark matter halo models for different sources, we  evaluate the integral in Eq.~\ref{eq:flux}, conventionally referred to as the D-factor, for each of the sources with $\Delta\Omega=2\pi\cos{\theta}$. Here, $\theta$ denotes the angular distance from the center of the source. Then, we select the targets with  larger D-factors than $10^{18}$ GeV/cm$^{2}$, which are listed in Table~\ref{table:1}. $\theta_{max}$ denotes the angular distance from the source at which the D-factor reaches its maximum, while $D_{max}$ is the D-factor calculated up to $\theta=\theta_{max}$. We use the CLUMPY software package~\cite{CLUMPY,CLUMPYv3} to calculate $D_{max}$ and the spatial distribution of the neutrino flux from the targets. The locations of the selected targets in the sky are depicted in Figure~\ref{fig:skymap}.

\begin{table} 
\centering
\resizebox{0.7\textwidth}{!}{
\begin{tabular}{|c|c|c|c|c|c|}
\hline
Source            & Type & RA[$^{\circ}$] & Dec [$^{\circ}$] & $\theta_{max}$ [$^{\circ}$] & $D_{max}$ [$GeV/cm^{2}$] \\
\hline
\hline
Virgo             & galaxy cluster & $186.63$  & $12.72$ & $6.11$ & $2.54\times10^{20}$    \\
Coma              & galaxy cluster & $194.95$  & $27.94$ & $1.30$ & $1.49\times10^{19}$    \\
Perseus           & galaxy cluster & $49.94$   & $41.51$ & $1.35$ & $1.44\times10^{19}$    \\
\hline
Andromeda         & galaxy         & $10.68$   & $41.27$ & $8.00$ & $1.70\times10^{20}$    \\
\hline
Draco             & dwarf galaxy & $260.05$ & $57.92$  & $1.30$ &  $1.54\times10^{19}$ \\
Leo I             & dwarf galaxy & $152.12$ & $12.3$   & $0.45$ &  $1.19\times10^{18}$  \\
Ursa Minor        & dwarf galaxy & $227.28$ & $67.23$  & $1.37$ &  $1.41\times10^{18}$  \\
Bo\"otes I        & dwarf galaxy & $210.03$ & $14.5$   & $0.47$ &  $1.46\times10^{18}$  \\
Coma Berenices    & dwarf galaxy & $186.74$ & $23.9$   & $0.31$ &  $1.53\times10^{18}$ \\
Segue 1           & dwarf galaxy & $151.77$ & $16.08$  & $0.35$ &  $2.05\times10^{18}$  \\
Ursa Major II     & dwarf galaxy & $132.87$ & $63.13$  & $0.53$ &  $5.23\times10^{18}$ \\ 
\hline
\end{tabular} }
\caption{Sources of neutrinos selected for this analysis. $\theta_{max}$ is the angular radius of the dark matter halo of the source at which the D-factor, the integral factor in Eq.~\ref{eq:flux}, reaches its maximum ($D_{max}$).} 
\label{table:1}
\end{table}     

\indent The expected background includes the neutrinos produced by cosmic-ray interactions with Earth's atmosphere and isotropic astrophysical neutrinos. The background is expected to be uniform in right ascension, as the detector rotates along the Earth's rotation axis. Thus, the scrambled data can be used to estimate the distribution of background events. This is done by replacing the right ascension of each event by a random value between $0$ and $2\pi$. 

\section{Data Sample and Statistical Methods} \label{sec:3}
For this work, we use a subset of the IceCube data sample established to search for point-like sources~\cite{IceCube:PSTracks_V3}. The sample contains track-like events from both the southern and northern sky. The data are collected between 2008 and 2018 of which we use the last 6 years of data. The data of the subset are taken with the full 86-strings detector configuration, for which an updated event selection is applied compared to previous years. With this track-like event sample, the detector has a median angular resolution smaller than $1^{\circ}$, for neutrino energies of $\sim$TeV or above. 

To determine whether there is a significant excess in data, we perform a hypothesis test using a likelihood ratio test statistic defined as below:
\begin{equation}
   \lambda = -2\log\frac{\mathcal{L}(n_{s}=0)}{\mathcal{L}(\hat{n}_{s})},
\end{equation}
where ${n_{s}}$ is the expected number of signal events, and $\hat{n}_{s}$ is the best-fit value of the parameter. The likelihood function is defined as the following: 
\begin{equation}
    \mathcal{L}(n_{s}) = \prod_{i=1}^{N} \left[\frac{n_{s}}{N}S(\alpha_{i}, \delta_{i}, \sigma_{i}, E_{i} | n_{s}) 
    + \left(1-\frac{n_{s}}{N}\right) B(\alpha_{i}, \delta_{i}, \sigma_{i}, E_{i} | n_{s})   \right],
\end{equation}
where $i$ is the event index, and $N$ is the total number of observed events. $\alpha$ and  $\delta$ are the reconstructed right ascension and declination, respectively. $\sigma$ is the angular uncertainty, and $E$ is the reconstructed neutrino energy. 

The sensitivity is defined, at 90\% confidence level (CL), as the minimum number of signal events required to have a Type I error rate smaller than 0.5 and Type II error rate of 0.1. From this, we can deduce the sensitivity on the dark matter decay lifetime using the following equation: 

\begin{equation}
   n_{s} = T_{\text{live}} \int_{0}^{\Delta \Omega} d\Omega \int_{E_{\text{min}}}^{E_{\text{max}}} dE_{\nu} A_{\text{eff}}(\hat{n}, E_{\nu}) \frac{d\Phi_{\nu}}{d\Omega dE_{\nu}}(\hat{n}, E_{\nu}), 
\end{equation}
\noindent where $T_{\text{live}}$ is the detector livetime, $A_{\text{eff}}$ is the effective area of the detector, and $E_{\text{min}}$, $E_{\text{max}}$ are the minimum, maximum energies of the expected neutrinos, respectively.

\section{Sensitivities} \label{sec:4}
\begin{figure}[t]
\centering
\includegraphics[width=.48\linewidth]{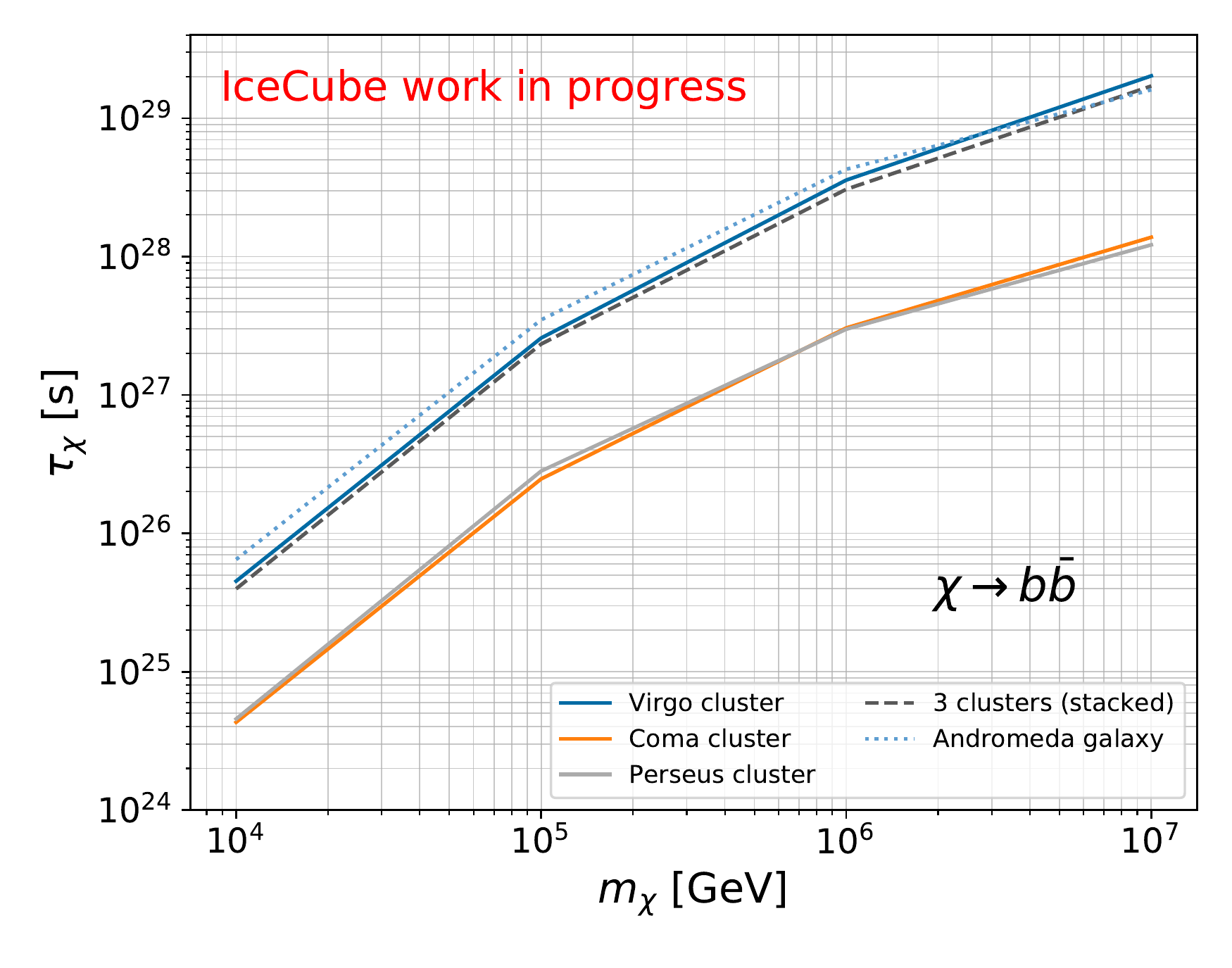} 
\includegraphics[width=.48\linewidth]{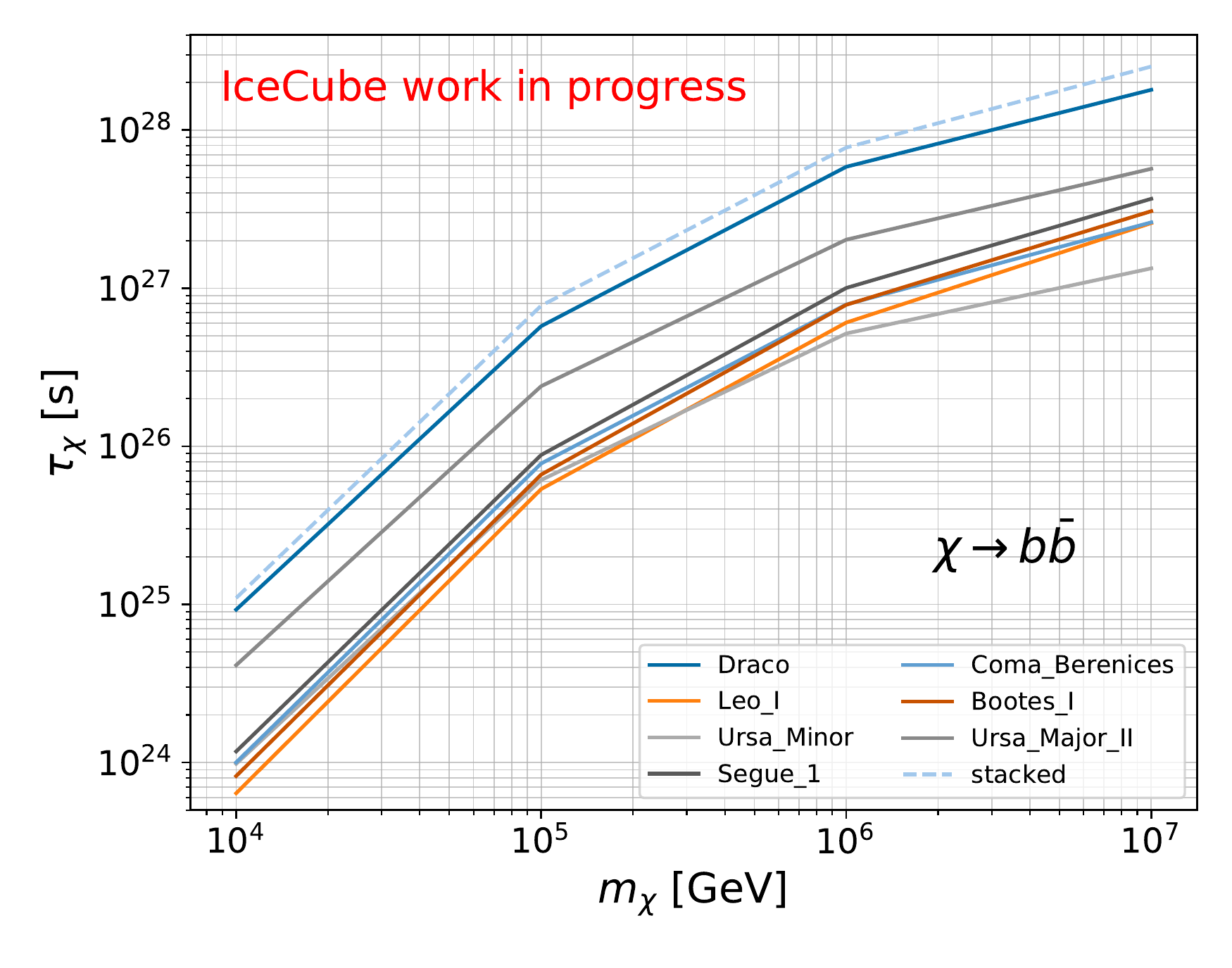} 
\caption[margin=1cm]{90\% CL sensitivities on dark matter lifetime for the $b\bar{b}$ channel. {\bf Left:} The sensitivities for the three galaxy clusters. The solid lines are for the individual galaxy clusters, and the dashed line is for the case of stacking the three clusters. The dotted line is for the Andromeda galaxy. {\bf Right:} The sensitivities for the seven dwarf galaxies. The solid lines are for the individual dwarf galaxies and the dashed line is for the case of stacking all of the dwarf galaxies.}
\label{fig:sens_PSTracks_bb}
\end{figure}

In this section, we present the 90\% CL sensitivities obtained for this  analysis. The sensitivities computed for dark matter decaying into $b\bar{b}$ can be seen in Figure~\ref{fig:sens_PSTracks_bb}. The left panel shows the sensitivities for the galaxy clusters and the Andromeda galaxy. The sensitivities for the Virgo cluster are better than those for the Coma cluster and Perseus cluster. This is mainly due to the fact that the D-factor for the Virgo cluster is much larger than those for the other two clusters. Stacking the three clusters does not improve the sensitivities, compared to using only the Virgo cluster. The sensitivities for the Andromeda galaxy are comparable to those for the Virgo cluster. In the right panel of Figure~\ref{fig:sens_PSTracks_bb}, we present the sensitivities for the seven dwarf galaxies that we selected. The sensitivities for the Draco galaxy and the Ursa Major II galaxy are better than those for the other dwarf galaxies. This can be explained by the relatively large D-factors for the two galaxies, similar to the case of the galaxy clusters. The sensitivities are improved by stacking all of the dwarf galaxies, but they are still not comparable to the sensitivities for the Virgo cluster and the Andromeda galaxy. 

\begin{figure}[b]
\centering
\includegraphics[width=.48\linewidth]{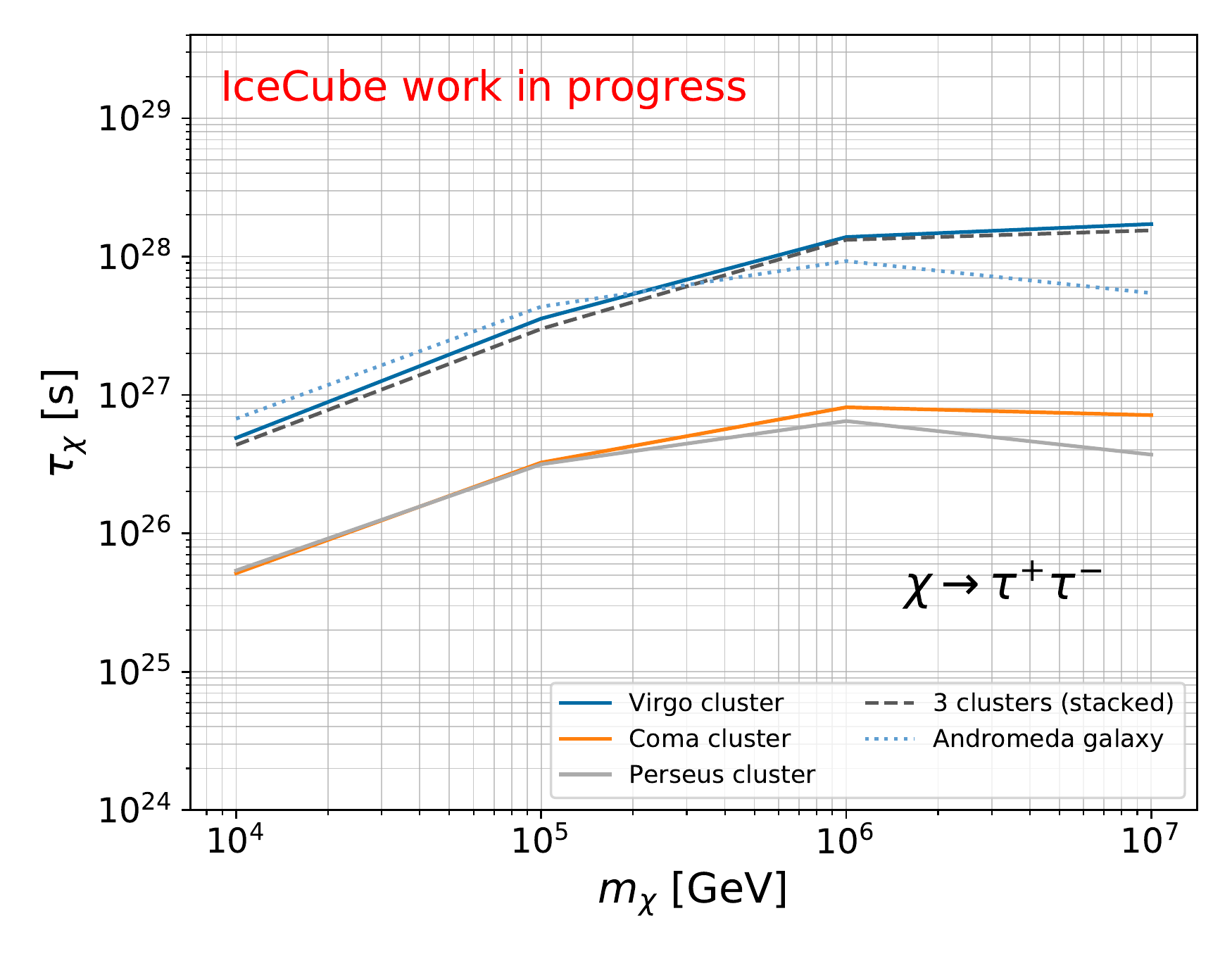} 
\includegraphics[width=.48\linewidth]{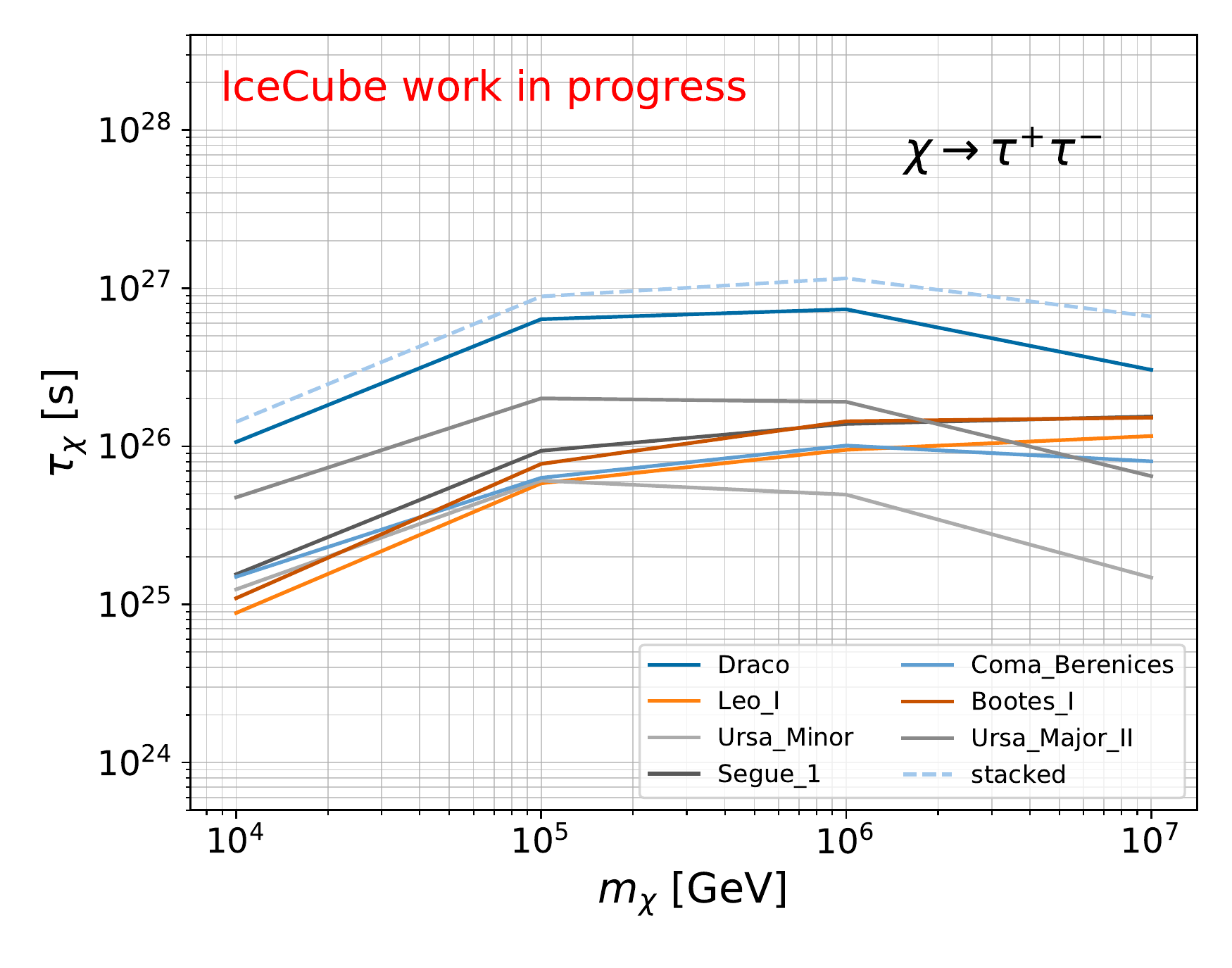} 
\caption[margin=1cm]{90\% CL sensitivities on dark matter lifetime for the $\tau^{+}\tau^{-}$ channel.}
\label{fig:sens_PSTracks_tautau}
\end{figure}

\begin{figure}[h]
\centering
\includegraphics[width=.54\linewidth]{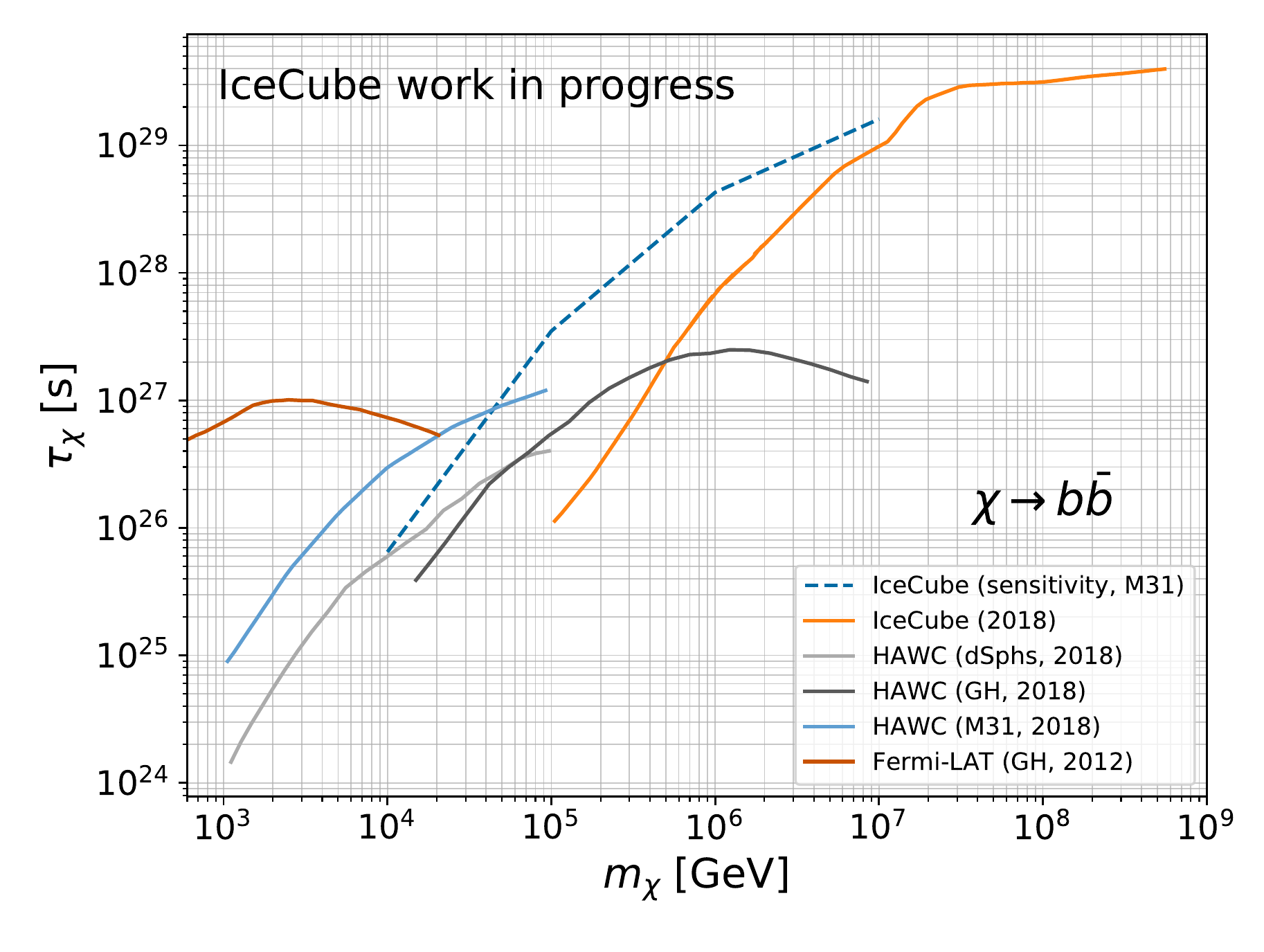} 
\caption[margin=1cm]{Comparison of the sensitivities of the analysis to the observed limits from other experiments~\cite{HAWC:M31, HAWC:dSphs, HAWC:GH, IceCube:GH, Fermi-LAT:GH}. The dashed line represents the 90\% CL sensitivities of the analysis for the $b\bar{b}$ channel and the Andromeda galaxy. Other lines are the limits from different experiments calculated assuming the same dark matter decay channel. The confidence levels associated with the IceCube limits, HAWC limits, and the Fermi-LAT limits are 90\%, 95\%, and 3$\sigma$, respectively.}
\label{fig:sens_comparison}
\end{figure}

\indent Figure~\ref{fig:sens_PSTracks_tautau} shows the sensitivities for the $\tau^{+}\tau^{-}$ channel. It can seen that for this channel the Virgo cluster and the Andromeda galaxy are the most promising sources. Stacking the dwarf galaxies leads to slightly better sensitivities than those for the best dwarf galaxy. The sensitivity curves for Perseus, Andromeda, Draco, Ursa Major II, and Ursa Minor have onsets of decrease. This is due to the opacity of the Earth to neutrinos. The opacity increases with increasing neutrino energy and increasing source declination. Thus, the Earth is more opaque to the neutrinos from these sources than the others. Furthermore, the higher the dark matter mass is, the larger is the opacity for these sources. Therefore, the sensitivities for the sources do not keep increasing with increasing dark matter mass. This effect is less noticeable  when dark matter decays into $b\bar{b}$, as the neutrino spectra for this channel are soft.  \newline
\indent We compare the sensitivities of this analysis to the observed limits from other experiments in Figure~\ref{fig:sens_comparison}. The dashed line represents the sensitivities of the presented analysis for the $b\bar{b}$ channel and the Andromeda galaxy. The solid lines are the limits from different experiments~\cite{HAWC:M31, HAWC:dSphs, HAWC:GH, IceCube:GH, Fermi-LAT:GH}, for the same dark matter decay channel. It can be seen that the sensitivities calculated for this analysis are competitive with other experiments. 

\section{Conclusions and Outlook} \label{sec:5}

We calculated the sensitivities, at 90\% CL, for the $b\bar{b}$, $\tau^{+}\tau^{-}$ channels for dark matter masses between 10\,TeV and 10\,PeV. For the Virgo cluster and the Andromeda galaxy, the sensitivities are competitive with other leading experiments. We also tested if the sensitivities could be improved by stacking the sources. For the $b\bar{b}$ and $\tau^{+}\tau^{-}$ channels, stacking the galaxy clusters did not improve the sensitivities, compared to using only the Virgo cluster. Furthermore, the sensitivities for the Virgo cluster and the Andromeda galaxy were significantly better than those obtained by stacking the dwarf galaxies. We plan to study the sensitivities for other dark matter decay channels than $b\bar{b}$ and $\tau^{+}\tau^{-}$. Galaxy clusters and dwarf galaxies in the southern sky could be included in the list of targets, and the benefit of stacking could be evaluated again.    

\bibliographystyle{ICRC}
\bibliography{refs}

\clearpage
\section*{Full Authors List: \Coll\ Collaboration}
%
%
%

\scriptsize
\noindent
R. Abbasi$^{17}$,
M. Ackermann$^{59}$,
J. Adams$^{18}$,
J. A. Aguilar$^{12}$,
M. Ahlers$^{22}$,
M. Ahrens$^{50}$,
C. Alispach$^{28}$,
A. A. Alves Jr.$^{31}$,
N. M. Amin$^{42}$,
R. An$^{14}$
% ...
\\

\noindent
% ...
$^{12}$ Universit{\'e} Libre de Bruxelles, Science Faculty CP230, B-1050 Brussels, Belgium \\
% ...
$^{14}$ Department of Physics and Laboratory for Particle Physics and Cosmology, Harvard University, Cambridge, MA 02138, USA \\
% ...
$^{17}$ Department of Physics, Loyola University Chicago, Chicago, IL 60660, USA \\
$^{18}$ Dept. of Physics and Astronomy, University of Canterbury, Private Bag 4800, Christchurch, New Zealand \\
% ...
$^{22}$ Niels Bohr Institute, University of Copenhagen, DK-2100 Copenhagen, Denmark \\
% ...
$^{28}$ D{\'e}partement de physique nucl{\'e}aire et corpusculaire, Universit{\'e} de Gen{\`e}ve, CH-1211 Gen{\`e}ve, Switzerland \\
% ...
$^{31}$ Karlsruhe Institute of Technology, Institute for Astroparticle Physics, D-76021 Karlsruhe, Germany  \\
% ...
$^{42}$ Bartol Research Institute and Dept. of Physics and Astronomy, University of Delaware, Newark, DE 19716, USA \\
% ...
$^{50}$ Oskar Klein Centre and Dept. of Physics, Stockholm University, SE-10691 Stockholm, Sweden \\
% ...
$^{59}$ DESY, D-15738 Zeuthen, Germany \\
% ...

\subsection*{Acknowledgements}

\noindent
USA {\textendash} U.S. National Science Foundation-Office of Polar Programs,
U.S. National Science Foundation-Physics Division,
U.S. National Science Foundation-EPSCoR,
Wisconsin Alumni Research Foundation,
Center for High Throughput Computing (CHTC) at the University of Wisconsin{\textendash}Madison,
Open Science Grid (OSG),
Extreme Science and Engineering Discovery Environment (XSEDE),
Frontera computing project at the Texas Advanced Computing Center,
U.S. Department of Energy-National Energy Research Scientific Computing Center,
Particle astrophysics research computing center at the University of Maryland,
Institute for Cyber-Enabled Research at Michigan State University,
and Astroparticle physics computational facility at Marquette University;
Belgium {\textendash} Funds for Scientific Research (FRS-FNRS and FWO),
FWO Odysseus and Big Science programmes,
and Belgian Federal Science Policy Office (Belspo);
Germany {\textendash} Bundesministerium f{\"u}r Bildung und Forschung (BMBF),
Deutsche Forschungsgemeinschaft (DFG),
Helmholtz Alliance for Astroparticle Physics (HAP),
Initiative and Networking Fund of the Helmholtz Association,
Deutsches Elektronen Synchrotron (DESY),
and High Performance Computing cluster of the RWTH Aachen;
Sweden {\textendash} Swedish Research Council,
Swedish Polar Research Secretariat,
Swedish National Infrastructure for Computing (SNIC),
and Knut and Alice Wallenberg Foundation;
Australia {\textendash} Australian Research Council;
Canada {\textendash} Natural Sciences and Engineering Research Council of Canada,
Calcul Qu{\'e}bec, Compute Ontario, Canada Foundation for Innovation, WestGrid, and Compute Canada;
Denmark {\textendash} Villum Fonden and Carlsberg Foundation;
New Zealand {\textendash} Marsden Fund;
Japan {\textendash} Japan Society for Promotion of Science (JSPS)
and Institute for Global Prominent Research (IGPR) of Chiba University;
Korea {\textendash} National Research Foundation of Korea (NRF);
Switzerland {\textendash} Swiss National Science Foundation (SNSF);
United Kingdom {\textendash} Department of Physics, University of Oxford.

\end{document}